# Controllable Andreev Bound States in Bilayer Graphene Josephson Junction from Short to Long Junction Limits

Geon-Hyoung Park[1], Wonjun Lee[1,2], Sein Park[1], Kenji Watanabe[3], Takashi Taniguchi[4],

Gil Young Cho[1,2,5] and Gil-Ho Lee[1,5*]

[1]Deparment of Physics, Pohang University of Science and Technology, Pohang, Republic of Korea

[2]Center for Artificial Low Dimensional Electronic Systems, Institute for Basic Science, Pohang, Republic of Korea

[3]Research Center for Electronic and Optical Materials, National Institute for Materials Science, 1-1 Namiki, Tsukuba 305-0044, Japan

[4]Research Center for Materials Nanoarchitectonics, National Institute for Materials Science, 1-1 Namiki, Tsukuba 305-0044, Japan

[5]Asia Pacific Center for Theoretical Physics, Pohang, Republic of Korea

**Abstract**

We demonstrate that the mode number of Andreev bound states in bilayer graphene Josephson junctions can be modulated by *in situ* control of the superconducting coherence length. By exploiting the quadratic band dispersion of bilayer graphene, we control the Fermi velocity and thus the coherence length by the application of the electrostatic gating. Tunneling spectroscopy of Andreev bound states reveals a crossover from short to long Josephson junction regimes as the gate voltage is approached near the charge neutral point of bilayer graphene. Furthermore, quantitative analysis of Andreev spectrums for different mode numbers allows us to quantitatively estimate the phase-dependent Josephson current. Our work paves a new way to study multi-mode Andreev levels and to engineer Fermi velocity with bilayer graphene.

In a superconductor-normal material-superconductor (SNS) Josephson junction (JJ), Cooper pairs can be transferred from one superconductor to another via the normal channel as coherently coupled electron and hole pairs generated by consecutive Andreev reflections at SN interfaces [1-3]. Andreev bound states (ABS) are formed as standing waves of Andreev pairs confined within a superconducting gap in the normal channel and play a key role in governing the physics of SNS junctions. Recently, there has been a growing interest in ABS in the research area of topological superconductivity [4-6], quantum information processing [7-9], quantum states manipulation [10-12], and application in quantum sensors [13,14]. Thanks to advanced nano-fabrication techniques, direct observation and manipulation of ABS have become feasible.

ABS comes in pairs at positive and negative energies due to the particle-hole symmetry of the superconductivity, and its energy exhibits periodic oscillation with the macroscopic superconducting phase difference between two superconductors of JJ. The number of ABS modes is determined by the ratio between superconducting coherence length ($\xi_0$) and channel length ($L$). For a ballistic conductor, superconducting coherence length is given by Fermi velocity $v_F$ and superconducting gap $\Delta$ as $\xi_0 = \hbar v_F/2\Delta$, and determines short ($L \ll \xi_0$) or long ($L \gg \xi_0$) junction limit [3,15]. Conventionally, the number of ABS pairs is one for short junctions, and it is two or more for long junctions. ABS in short junctions have been observed by tunneling spectroscopy in various systems such as two-dimensional electron gas (2DEG) system [16], carbon nanotube [17], semiconducting nanowires [18-21], and graphene [22-25], while only a few studies have reported the observation of ABS in long junctions [26,27]. Moreover, there have been no attempts to systematically verify a relationship between the number of ABS pairs and the ratio $L/\xi_0$. To study more details of ABS in different junction limits, we propose a new system that allows *in situ* control of $\xi_0$ while simultaneously observing ABS.

In this Letter, we adopted bilayer graphene (BLG) as a normal weak link in SNS JJ to study the crossover between short and long junction limits by controlling $\xi_0$. The low-energy bands of BLG exhibit approximately quadratic dispersion $E \approx p^2/2m_e^*$ at small momentum [28] so the Fermi velocity is given by $v_F = \frac{1}{\hbar}\frac{\partial E}{\partial k_F} = \frac{\hbar\sqrt{\pi n}}{m_e^*}$, where $m_e^*$ is the effective mass of electron and *n* is the carrier density. In this work, we used *in situ* electrostatic gating to control $v_F$, hence $\xi_0$. For observing ABS formed in the BLG channel, tunneling spectroscopy with an edge-contacted superconducting tunneling probe is performed [24]. We confirmed the number of ABS pairs varied from one up to three by applying the backgate voltage to the device. We were also able to construct the current-phase relation of ABS without extra DC SQUID measurement. In addition, important parameters such as $v_F$ and $\tau$ of the JJ were estimated by fitting theoretical curves to the ABS energy.

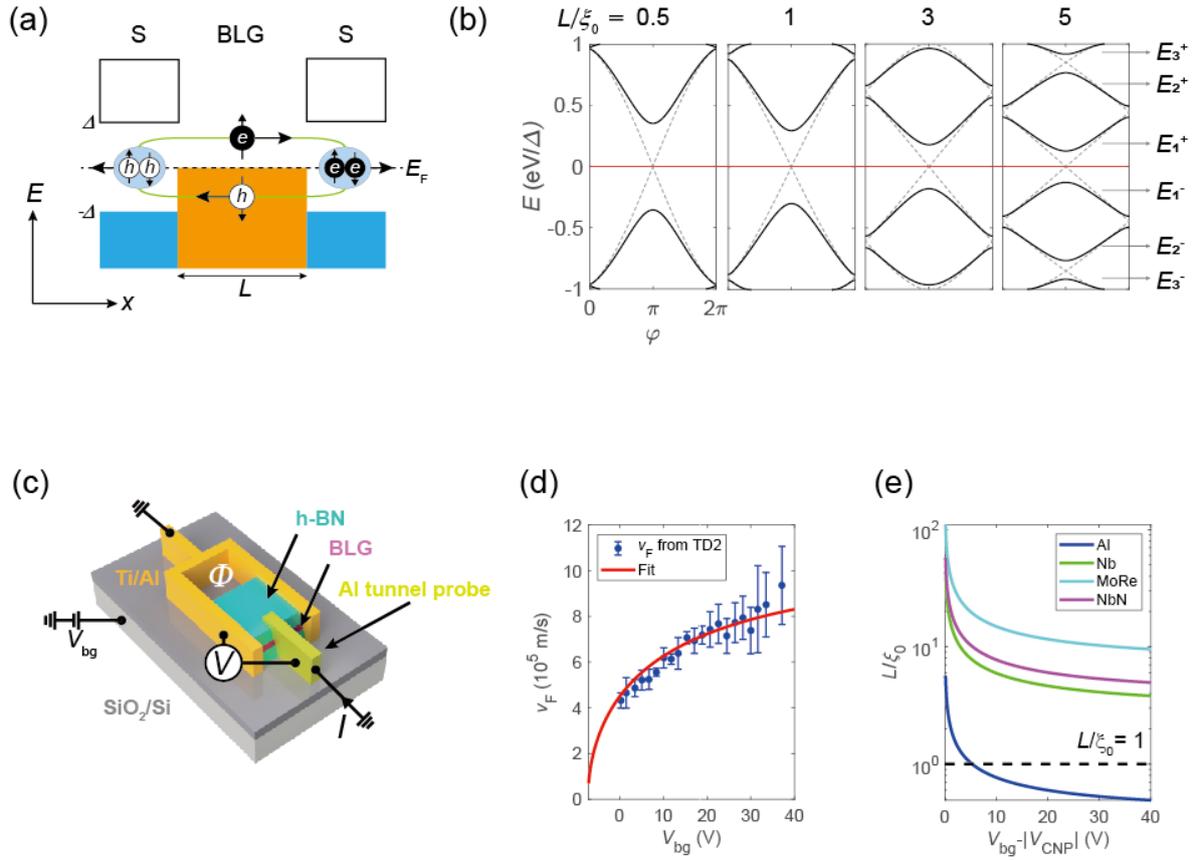

**Figure 1** (a) A schematics of successive Andreev reflections in superconductor-normal metal-superconductor (SNS) junction. (b) Energy-phase relationship of Andreev bound states (ABS) with different ratios between junction length $L$ and superconducting coherence length $\xi_0$, $L/\xi_0$ values. Solid lines and dashed lines correspond to the transparency $\tau = 0.9$ and 1, respectively. (c) A schematic of the device LD2 with the measurement configuration for tunnel spectroscopy. (d) Backgate voltage ($V_{bg}$) dependence of Fermi velocity estimated from Shubnikov-de Haas oscillation from the device TD2. (e) Backgate voltage dependence of $L/\xi_0$ calculated for different superconducting materials with Fermi velocity from (d) and $L = 1$ μm.

A microscopic mechanism of Josephson coupling is that electrons and Andreev-reflected holes go through consecutive Andreev reflections at the superconducting contacts and convey Cooper pairs from one superconductor to another (Fig. 1(a)). This roundtrip of quasiparticles results in bound states, which is called ABS. The ABS follows the Bohr-Sommerfeld quantization rule as [2].

$$2\cos^{-1}\left(\frac{E}{\Delta}\right) + \left(\frac{L}{\xi_0}\right)\left(\frac{E}{\Delta}\right) \pm \varphi = 2\pi N,$$

where $\varphi = \varphi_L - \varphi_R$ is the superconducting phase difference between left (L) and right (R) superconductors, and $N$ is an integer. In a short junction limit ($L \ll \xi_0$), the energy-phase relationship of ABS is simply described as $E_1^\pm = \pm\Delta\sqrt{1 - \tau \sin^2(\varphi/2)}$ with transparency $\tau$ [29]. In a long junction limit, however, there is no analytic expression for the ABS energy. Although there are some experimental reports on ABS in long junction limits [26,27] and a few theoretical studies on analytical approaches for ABS in the imperfect channel [30,31], experimental studies on ABS in long junction limits with quantitative analysis have not been carried out yet.

We fabricated two types of JJs; one with a superconducting loop for biasing the superconducting phase difference (LD2) and another for biasing current through the device (TD2) (see Fig. S1 in Supplemental Material). First, we encapsulate BLG with two distinct hexagonal boron nitride (*h*-BN) sheets to protect BLG from any chemical impurities during nanofabrication processes [32]. We performed atomic force microscopy on the BLG heterostructure and selected a flat area to fabricate the device on. 70 nm-thick Al superconducting electrodes together with a 6 nm-thick Ti adhesion layer are deposited onto the freshly etched one-dimensional edge of graphene [32,33] to induce Josephson coupling via BLG (yellow in Fig. 1c). Lastly, superconducting side tunnel contacts are made with deposition of 70 nm-thick Al electrodes on the edge of BLG. When Al directly contacts graphene, a potential barrier is expected to be formed between Al and graphene [34] so that only quantum tunneling can occur at a low temperature. This idea was adopted in our previous work to measure ABS in monolayer graphene-based JJs [24]. Tunneling conductance is measured by biasing the current $I$ to the tunneling probe while the superconducting loop is grounded and measuring the voltage $V$ between the tunneling probe and superconducting loop as depicted in Fig. 1(c). The magnetic flux $\Phi$ threading the superconducting loop is applied by external perpendicular magnetic field $B$. The backgate voltage $V_{bg}$ is applied to BLG via 300-nm-thick $SiO_2$ dielectric layer and 26-nm-thick bottom *h*-BN layer to modulate the carrier density and Fermi velocity of quasiparticles in BLG. All the data were measured at the base temperature of 17 mK, except for the *T* dependence data.

The modulation of Fermi velocity was estimated by analyzing the temperature dependence of Shubnikov-de Haas oscillation (SdHO) measured at the device TD2 (Fig. S4 in Supplement Material). Device TD2 shares the same BLG sheet with device LD2 with the same junction geometry. The amplitudes of SdHO are fitted to the Liftshitz-Kosevich formula [35,36] from the temperature $T = 1.5$ to 50 K. From this analysis, the effective mass $m_e^*$ of quasiparticles in BLG was estimated at different $V_{bg}$. Figure 1(d) shows the Fermi velocity estimated as a function of $V_{bg}$ using the relation $v_F = \hbar k_F/m_e^*$ from quadratic energy dispersion of BLG. The charge neutral point (CNP) at $V_{bg} = -7.2$ V is determined by measuring the two-terminal resistance as a function of the backgate voltage of device TD2, which is also adopted for the CNP of device LD2. We found that the experimental data is well-fitted to the theoretical model of Fermi velocity based on the well-known band dispersion of BLG without considering the interlayer asymmetry [28,37]. (see Fig. S3-4 in Supplemental Material for details). With estimated $v_F$, we also calculated $L/\xi_0$ as a function of $V_{bg}$ assuming various kind of superconductors with $L = 1$ μm [Fig. 1(e)]. As BLG is in a ballistic limit away from CNP, we used the relation of $\xi_0 = \hbar v_F/2\Delta$ (refer to Supplemental Material for discussion for ballistic transport of the BLG channel). Considering the maximum $V_{bg}$ before dielectric breakdown of SiO$_2$ layer is around 100 V, Al has the right size of superconducting gap ($\Delta = 129$ μeV) to observe the crossover from short to long junction limit.

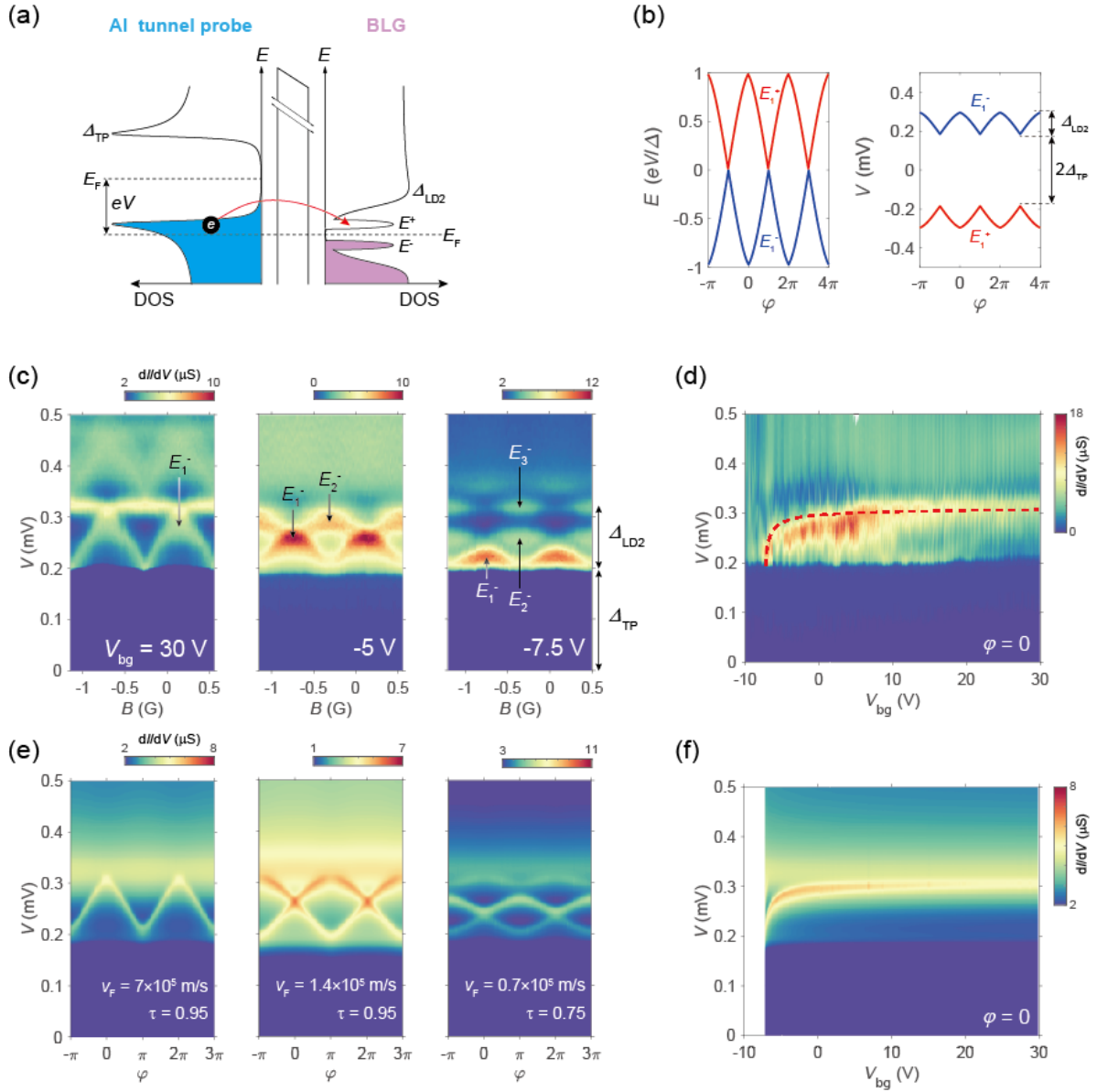

**Figure 2** (a) Elastic tunneling process of an electron from superconducting tunneling probe to Andreev bound state (ABS) in bilayer graphene (BLG) when tunneling differential conductance (d$I$/d$V$) peak is expected ($eV = \Delta_{TP} + E^+$). (b) (Left panel) Phase ($\varphi$) dependence of upper (red line) and lower ABS energy (blue line) in a short junction limit with perfect transparency. (Right panel) Expected tunneling conductance peak voltage $V$ as a function of $\varphi$. (c) Color plots of d$I$/d$V$ as a function of bias voltage $V$ and magnetic field $B$ for different backgate voltages $V_{bg}$ measured in device LD2. (d) A colormap of backgate voltage dependence of tunnel differential conductance (d$I$/d$V$) as a function of bias voltage ($V$) at a fixed phase difference $\varphi = 0$. The red dashed curve represents theoretically calculated $E_1^-$ from (f). (e-f) Theoretical simulations for (c-d) with $L/\xi_0$ obtained from device TD2, respectively.

Tunneling spectroscopy has recently been applied to graphene-based devices with normal [22,23] and superconducting tunneling probes [16,17,24,38]. Here, we performed tunneling spectroscopy with a superconducting tunneling probe onto BLG embedded into a Josephson junction to study the phase dependence of ABS. The superconducting tunneling probe allows us for higher energy resolution of spectroscopy than using a normal probe due to the sharp density of states (DOS) peaks near superconducting gap edges. With the biased energy $eV = \Delta_{TP} + E^+$ as depicted in Fig. 2(a), the filled DOS peak of probe (denoted by $\Delta_{TP}$) aligns to the empty DOS peak of upper ABS (denoted by $E^+$) where tunneling differential conductance (d$I$/d$V$) peak is expected. Tunneling differential conductance, however, does not simply represent the DOS of ABS, but it is determined by the convolution of the DOS of the probe and that of the sample [16,17,24,25]. Figure 2(b) shows the phase dependence of ABS energy and expected d$I$/d$V$ peak positions that are offset by the superconducting gap of the tunneling probe. Figure 2(c) shows colormaps of d$I$/d$V$ as a function of $V$ and $\varphi$ at $V_{bg}$= 30, -5, and -7.5 V (left to right). The amplitude of the modulation of d$I$/d$V$ peaks with $\varphi$ corresponds to $\Delta_{LD2} = 129$ μeV and the offset of the modulation corresponds to $\Delta_{TP} = 185$ μeV. Superconducting phase difference $\varphi = 2\pi\Phi/\Phi_0 = 2\pi(B - B_0)A/\Phi_0$ is controlled by external magnetic field $B$, with the loop area $A$ and the magnetic flux quantum $\Phi_0$. The offset of magnetic field $B_0$ = -0.72 G is manually determined by considering the center of the Fraunhofer diffraction pattern of the device TD2 (Fig.S2 in the Supplemental Material). The ABS oscillation period is 0.86 G, which is consistent with $\Phi_0/A = 0.85$ with $A \sim 24\ \mu m^2$.

By tuning gate voltages, we observed the crossover from short to long junction limit. At $V_{bg}$ = 30 V, there is only one ABS pair ($N$ = 1) oscillating within the gap $\Delta_{LD2}$, which indicates that the junction is in the short junction limit. This is consistent with the expectation with $L/\xi_0 \sim 0.5$ being smaller than 1. As the gate voltage approaches the CNP, the second ($N = 2$) and third ABS pairs ($N = 3$) gradually appear and the amplitude of d$I$/d$V$ peak oscillation decreases. At $V_{bg}$ = -5 V, the second ABS ($E_2^-$) that oscillates out-of-phase with the first ABS ($E_1^-$) is clearly seen. Near CNP ($V_{bg}$ = -7.5 V), the third ABS ($E_3^-$) starts to appear and the oscillation amplitude becomes very small. We could not observe ABS pairs with $N \geq 4$ due to limited energy resolution. Nonetheless, adopting larger gap superconductors such as TaN (0.7 meV) [39], Nb (1 meV) [40-42], and MoRe (1.4 meV) [33,43] might be a viable approach to study higher modes of ABS. We also observed additional d$I$/d$V$ peak oscillations at $eV > |\Delta_{TP} + \Delta_{LD2}|$ for highly n-doped region around $V_{bg}$ = 15 ~ 30 V (see the Supplemental Material for other backgate voltages). These peaks mimic the original ABS oscillations and appear repeatedly up to $V = \pm 1.2$ meV with specific energy spacings $\sim 0.12 - 0.36$ meV. We suspect that they may originate from the multiple reflections at contacts or edges in the devices as discussed in other studies [44-46], or from the inelastic tunneling process due to impurity states in the tunneling barrier [23]. We also measured the gate voltage dependence of d$I$/d$V$ at a fixed phase difference $\varphi = 0$ to confirm the variance of $E_1^-$ more clearly as shown in Fig. 2(d). We observed the positions of a d$I$/d$V$ peak ($E_1^-$) gradually decrease as $V_{bg}$ approaches the CNP. The decreasing behavior becomes noticeable at $V_{bg} < 5$ V, indicating the increase in $L/\xi_0$ beyond unity.

For more quantitative analysis, we introduced a theoretical model for ABS considering the geometrical asymmetry of the tunnel contact as an effective scatterer (see the Supplemental Material for a detailed description). The numerical simulation on ABS (Figs. 2(e) and (f)) by using $v_F$ value obtained in the device TD2 (red solid line in Fig. 1(e)), and the transparency $\tau = 0.75 \sim 0.95$ successfully demonstrates the decrease in ABS energy and the reduction in oscillation amplitude near the CNP. Our simulation results also explain the energy gaps between $E_2^-$ and $E_3^-$ at $\varphi = \pi n$ in Figs. 2(c) and (e), which occurs due to the low transparency near the CNP [30,31]. However, d$I$/d$V$ peaks at $V_{bg} < 10$ V are located at slightly lower bias voltage and have broader peaks than the theoretical ones. This can be understood by the mean free path ($l_m$) of BLG becoming shorter than the channel length as approaching the CNP. This demands the superconducting coherence length in a diffusive limit $\xi_0 = \sqrt{\hbar D/\Delta}$ much shorter than $\xi_0$ in a ballistic limit, resulting in the increase of $L/\xi_0$. Here, $D = v_F l_m/2$ is Einstein diffusion coefficient.

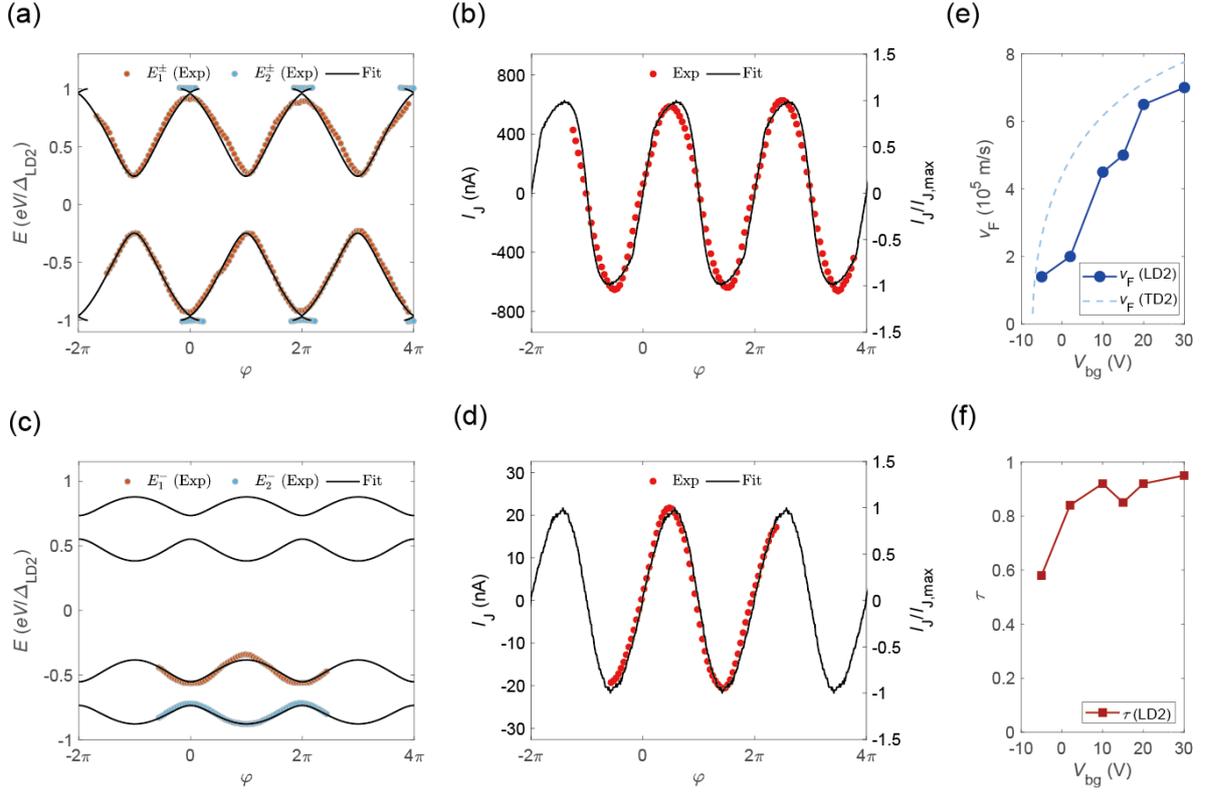

**Figure 3** Current-phase relationship in short and long junction limit. (a), (c) Phase ($\varphi$) dependence of ABS energy extracted from the d$I$/d$V$ peaks at $V_{bg}$ = 30 and -5 V, respectively. Solid black curves are theoretical fits to the experimental data (symbols). (b), (d) Current-phase relations calculated from the ABS energy in (a) and (c), respectively, and the theoretical fits (solid black curves). (e) Fermi velocity in devices LD2 and TD2 as a function of backgate voltage ($V_{bg}$). (f) Transparency $\tau$ of LD2 as a function of $V_{bg}$.

Current-phase relationship (CPR) can reveal more abundant characteristics of the Josephson supercurrent. In a zero-temperature limit, the Josephson current can be obtained by the summation of contributions from $N$th ABS that are filled below the Fermi level, $I_J = \sum_N I_N$, where $I_N = -\frac{2e}{h}\frac{\partial E_N^-}{\partial \varphi}$. From this, we can estimate the Josephson current of the junction by extracting $E_N^-$ from the data in short ($N = 1$) and long ($N \geq 2$) junction limits. We obtained ABS energies by using an approximation $E_N^- \sim -(eV_{ABS,N}^- - \Delta_{TP})$, instead of performing a deconvolution of the measured tunneling conductance. Here, $eV_{ABS,n}^-$ represents the bias energy at which the tunneling conductance peak occurs by the ABS below the Fermi level. Figure 3(a) shows extracted $E_N^\pm$ at $V_{bg}$ = 30 V and their theoretical fits with $L/\xi_0 = 0.56$ and $\tau = 0.95$. The CPR data of Fig. 3(a) also demonstrates a good agreement between experiment and theory [Fig. 3(b)]. The oscillation of $I_J$ mainly follows $E_1^-$ since $E_2^-$ is not dominant here ($I_J \sim I_1$). From the $L/\xi_0$ and $\tau$, we can deduce that the LD2 device at $V_{bg}$ = 30 V is indeed in the short junction and ballistic limit. On the other hand, Figs. 3(c) and (d) show the case when $E_2^-$ is also dominant ($V_{bg}$ = -5 V). The theoretical calculations fit well with the data with $L/\xi_0 = 2.79$ and $\tau = 0.58$, indicating that the junction is in the long junction. In Fig. 3(d), $I_1$ can be mostly canceled out by $I_2$ which oscillates in antiphase to $I_1$, resulting in suppressed $I_J$. It indicates that the total Josephson current decreases as $N$ increases, and it also oscillates when $N$ is even or odd. To quantify the $I_J$ in Figs. 3(b) and (d), the number of conduction channels in BLG $M \sim 2W/\lambda_F$ is also considered, where $W$ is the width of the channel and $\lambda_F$ is the Fermi wavelength of electrons. $M$ can vary from $\sim$ 60 to 260 with $V_{bg}$ ranging from -5 to 30V. At $V_{bg}$ = -5 V, the full participation of $E_2^-$ strongly suppresses the total Josephson current, even considering the fourfold reduction of the channel number.

Until now, analyzing CPR in monolayer graphene and 2DEG-based JJs has been significant in verifying the $\tau$ of the junction [16,22]. However, in our BLG JJ, the Fermi velocity also holds significance as a fitting parameter, allowing us to reconfirm the variation of the $\xi_0$ in LD2 device. The comparison of two Fermi velocities estimated for each TD2 (blue dashed curve) and LD2 (blue dots) is shown in Fig. 3(e). It can be observed that the estimated $v_F$ for both devices shows a similar trend depending on the $V_{bg}$. In Fig. 3(f), the transparency $\tau$ in LD2 also has a strong dependence on gate voltage especially near the CNP, suggesting that Andreev pairs might be affected by inhomogeneous electron-hole puddles in the BLG channels as we discussed in Fig. 2.

In conclusion, we achieved *in situ* control of the ABS number by exploiting the parabolic energy bands of BLG as a weak link in the JJ. As modulating the carrier concentration from far to near the CNP, $L/\xi_0$ was successfully varied from 0.5 to 5 without changing the channel length or replacing the superconducting material. The gate dependence of tunneling conductance shows a crossover from the short-like ($N \sim 1$) to the long-like ($N \sim 3$) junction in LD2 near the CNP of BLG. Moreover, inspecting current-phase relations for short and long-like junction limits reveals that the even pairs of ABS strongly suppress the Josephson current and the precise Fermi velocity and transparency values can be extracted. We conducted the first full-scale experimental analysis of ABS in the long junction limit and demonstrated that the ABS number $N$ varies according to the relationship between the channel length and the superconducting coherence length. We expect that this work will give new possibilities in Andreev-multi-level physics and Fermi velocity engineering with bilayer graphene.

**Acknowledgments**

This work was supported by National Research Foundation (NRF) grants (Nos. 2021R1A6A1A10042944, 2022M3H4A1A04074153, RS-2023-00207732, RS-2023-00208291, No. 2023M3K5A1094810, No. 2023M3K5A1094813) and ITRC program (IITP-2022-RS-2022-00164799) funded by the Ministry of Science and ICT, the Air Force Office of Scientific Research under Award

No. FA2386-22-1-4061, Institute of Basic Science under project code IBS-R014-D1, Samsung Science and Technology Foundation under Project Number SSTF-BA2002-05, and Samsung Electronics Co., Ltd. (IO201207-07801-01). G.-H. P. was supported by the Basic Science Research Institute Fund (Grant No. 2021R1A6A1A10042944). K.W. and T.T. acknowledge support from the JSPS KAKENHI (Grant Numbers 21H05233 and 23H02052) and World Premier International Research Center Initiative (WPI), MEXT, Japan.

**Authors' e-mail**

*To whom all correspondence should be addressed: lghman@postech.ac.kr (G.-H.L.)